\documentclass[doublecol]{epl2}

\title{Geometric magnetoconductance dips by edge roughness in graphene nanoribbons}
\shorttitle{Title} 

\author{Hengyi Xu\inst{1} \and T. Heinzel\inst{1} \and I. V. Zozoulenko\inst{2}}
\shortauthor{F. Author \etal}

\institute{
  \inst{1} Condensed Matter Physics Laboratory, Heinrich-Heine-Universit\"at,
Universit\"atsstr.1, 40225 D\"usseldorf, Germany\\
  \inst{2} Solid State Electronics, Department of Science and Technology, Link\"{o}ping
University, 60174 Norrk\"{o}ping, Sweden
}

\abstract{The magnetoconductance of graphene nanoribbons with rough zigzag and
armchair edges is studied by numerical simulations. nanoribbons with
sufficiently small bulk disorder show a pronounced magnetoconductance minimum at cyclotron
radii close to the ribbon width, in close analogy to the \emph{wire peak} observed in
conventional semiconductor quantum wires. In zigzag nanoribbons, this feature becomes visible only above a threshold amplitude of the edge roughness, as a consequence of the reduced current density close to the edges.}

\begin{document}

\maketitle

Graphene, i.e. a single layer of graphite, possesses an energy dispersion
close to the charge neutrality point which is equivalent to the Dirac
dispersion for massless fermions.\cite{Novoselov2005,CastroNeto2009} This
property has raised tremendous activities in fundamental research in the
past few years, while the high room temperature electron mobility \cite%
{Morozov2008} of up to $20,000\,\mathrm{cm}^2/\mathrm{Vs}$ together with
excellent mechanical properties has raised hopes for novel electronics
applications. One key element of graphene nanostructures and devices are
ribbons with submicron width $W$, also known as \emph{graphene nanoribbons
(GNRs)}. It is well established that in GNRs, edge disorder can contribute
significantly to the scattering \cite%
{Areshkin2007,Evaldsson2008,Xu2008,Han2007} which is governed in wide
structures by a combination of scattering at charged impurities and resonant
scattering at short-range defects. \cite{Peres2011,Sarma2011} Furthermore,
edge roughness has been suggested as the source of the transport gap in
narrow GNRs around the charge neutrality point.\cite%
{Han2007,Stampfer2009,Liu2009,Todd2009} It would therefore be desirable to
be able to characterize and quantify the edge roughness in GNRs via
transport experiments. Conventional quantum wires made from, e.g., III/V
semiconductors, show a characteristic magnetoresistance maximum at a
perpendicular magnetic field which corresponds to a cyclotron radius $%
r_c\approx 0.55 W$, commonly refereed to as \emph{wire peak},\cite%
{Thornton1989} which can be interpreted as an enhanced diffusive scattering
when the classical electron trajectory grazes at the wire edge. \cite%
{Thornton1989,Akera1991} Within a quantum picture, the reduced conductance
emerges from a homogenized contribution of the occupied wire modes to the
overall conductivity. \cite{Akera1991} Since the strength of this effect
depends on the parameters that characterize the edge roughness, namely its
amplitude and correlation length which in turn determine the specularity of
the edge scattering, \cite{Thornton1989,Akera1991} the wire peak is
routinely used to characterize the edge roughness. \cite%
{Thornton1989,Lettau1994,Schapers2006,Gilbertson2011} We are not aware of
any observation of a similar structure in GNRs, where magnetoresistance
peaks in the parameter range of interest are absent. \cite%
{Poumirol2010,Bai2010,Oostinga2010,Ribeiro2011} This difference between GNRs
and conventional quantum wires has remained unexplained. The goal of the
present paper is to fill this gap and discuss theoretically the edge
roughness induced magnetoconductance dip (ERID) in GNRs. Our results show
that this structure should be observable in GNRs as well provided the
bulk disorder is sufficiently low. Furthermore, in zigzag GNRs, the structure is strongly suppressed at small edge disorder amplitudes.

\begin{figure}[tbp]
\includegraphics[width=85mm]{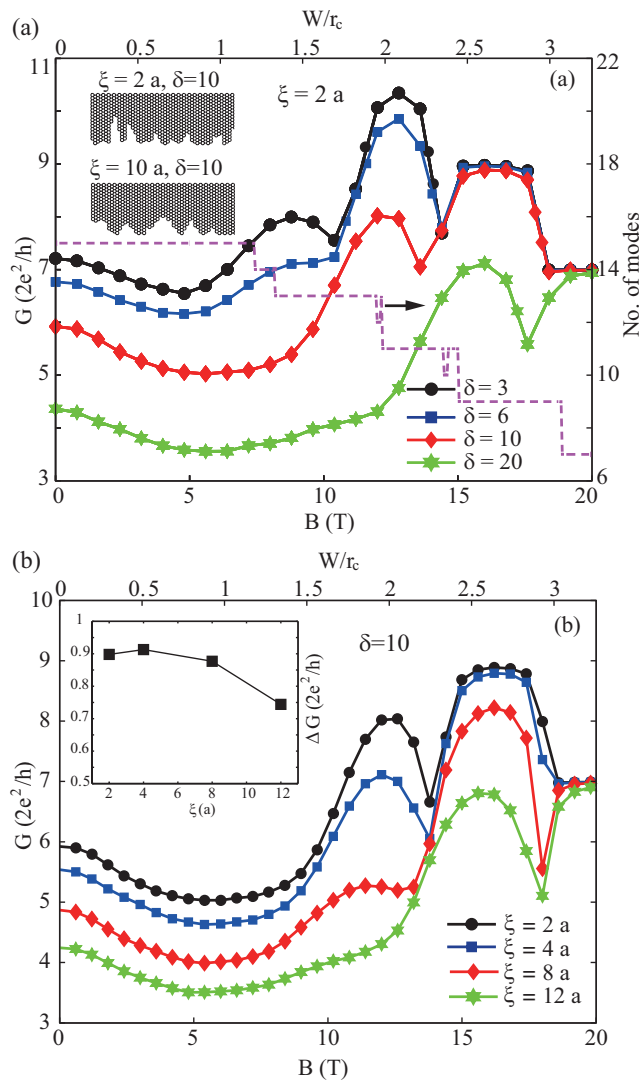}
\caption{(Colour online) Magnetoconductance of armchair GNRs (width $50\,\mathrm{nm}$,
length $220\,\mathrm{nm}$) as a function of the edge roughness parameters $%
\protect\delta$ (a) and $\protect\xi$ (b). Each trace is averaged over 500
edge configurations. The dashed lines show the number of occupied
subbands. Inset in (a): typical edge profiles for specific parameter values.
Inset in (b): amplitude of the ERID as a function of $\protect\xi$.}
\label{ERID_Fig1}
\end{figure}

We start with perfect armchair or zigzag nanoribbons oriented in x-direction
and of width $W$ in y-direction. Edge roughness is simulated at both edges
by missing atoms. It is parameterized by the maximum number $\delta$ of
missing atoms in y-direction from the edges into the GNR, and by the
correlation length $\xi$. The positions of the notches in x-direction are
chosen randomly with a probability $p$, while their individual central depth
$d^c$ is distributed equally between $0$ and $\delta$. Then the atoms around
the defect centers are removed following the Gaussian-type shape such that
the superimposed notches produce the number of missing atoms at edge site $%
x_j$
\begin{equation}
d (x_j)=\sum_{i=1}^{N_{c}}d^c_i\exp \left( -\frac{|x_i-x_j|^{2}}{2\xi^2}%
\right)  \label{Gaussian1}
\end{equation}
Here, $N_c=p\cdot N_{atom}$ is the number of notch centers with $N_{atom}$
being the number of carbon atoms along the edges. The roughness patterns on
both edges are uncorrelated. Typical structures are shown for one edge in
the inset of fig. \ref{ERID_Fig1} (a).

Besides the edge roughness, we also study the ERID effect under the influence of bulk disorder 
{which is modeled by the widely used random Gaussian potential \cite{Bardarson2007,Lewenkopf2008}
\begin{equation}
V_{i}=\sum_{i^{\prime }=1}^{N_{imp}}U_{i^{\prime }}\exp \left( -\frac{|%
\mathbf{r}{_{i}}-\mathbf{r}_{i^{\prime }}|^{2}}{2\Lambda ^{2}}\right)
\label{Gaussian2}
\end{equation}
The potential heights $U_i$ are uniformly distributed in the range
$U_{i}\in \lbrack -\Delta ,\Delta ]$ where $\Delta $ denotes the maximum
potential height. The potential decay depends on the site-site distance $|\mathbf{r}_i-\mathbf{r}_j|$ and the correlation length $\Lambda$. We refer to the case $\Lambda=a$ (appropriate for the absorbed neutral impurities) as
short-range scattering, and to the case $\Lambda \gg a$ (appropriate for the
remote charged impurities) as long-range scattering, with $a=0.142\,\mathrm{nm}$ being
the C-C bond length.}

Our calculations are based on the tight-binding Hamiltonian of GNRs
\begin{equation}
H=\sum_{i\neq j} t_{ij}c_i^\dag c_j+\sum_i V_{i}c_i^\dag c_i
\end{equation}
with the hopping energy $t_{ij}=t=-2.7\,\mathrm{eV}$ between the nearest
neighboring carbon atoms at zero magnetic field. The perpendicular magnetic
field is incorporated via Peierls' substitution $t_{ij}\rightarrow
t_{ij}e^{ie/\hbar \int_i^j \mathbf{A}\cdot d\mathbf{l}}$ with the vector
potential $\mathbf{A}=(-By,0,0)$ in the Landau gauge. 

The conductance of GNRs is calculated with Landauer-B\"{u}ttiker formula $%
G=2e^2/hT(E_F)$, where the transmission $T(E_F)$ is computed with the aid of the
recursive Green's function technique \cite{Xu2008}. Throughout this work,
the Fermi energy $E_F=0.1t$ corresponding to an electron density of $%
n=5.3\times 10^{16}\,\mathrm{m}^{-2}$ is used. The Fermi wavelength $%
\lambda_F$ is obtained from the linear band relation $E_F=\hbar v_Fk_F$ and $%
\lambda_F=2\pi/k_F$ with the Fermi velocity $v_F=3ta/2\hbar$. The cyclotron
radius is given by $r_c=\hbar k_F/eB$. 

Fig. \ref{ERID_Fig1} shows the results for the magnetoconductance $G(B)$ of
armchair GNRs of width $W=50\,\mathrm{nm}$ and length $L=220\,\mathrm{nm}$
for various values of the edge roughness parameters and in the absence of bulk
disorder. As the magnetic field is increased from zero, $G(B)$ decreases and
forms a broad minimum, the ERID, at $B_{min}\approx 5\,\mathrm{T}$
corresponding to $W/r_c\approx 0.8$. By varying W in the simulations, we find that $B_{min}$ is nearly inversely
proportional to the GNR width $W$ (not shown). This structure is the analogue of the
so-called \emph{wire peak} observed in conventional semiconductor quantum
wires. \cite{Thornton1989} The amplitude of the ERID depends only weakly on
the edge roughness amplitude $\delta$, Fig. \ref{ERID_Fig1}(a), and
decreases as the correlation length of the roughness $\xi$ is increased,
Fig. \ref{ERID_Fig1}(b). Outside the dip, a substantial increase of
the conductance follows as $B$ is increased, corresponding to the positive magnetoconductance
observed in recent experiments. \cite{Poumirol2010,Oostinga2010} As the
magnetic field is increased further beyond $\approx 12\,\mathrm{T}$,
oscillatory structures appear which reflect the magnetodepletion of the GNR
modes and are of no further interest here. We note that the effective widths of
GNRs are reduced by rough edges such that their signature in the
magnetoconductance is slightly displaced with respect to the
magnetodepletion of modes in defect-free GNRs. The ERID occurs in an
interval where the number of occupied modes does not change much, which
indicates that the structure is not related to the diamagnetic shift of the
modes. This behavior is in qualitative agreement with the results of the
quantum treatment of the wire peak in conventional quantum wires by Akera
and Ando.\cite{Akera1991}

\begin{figure}[tbp]
\includegraphics[width=85mm]{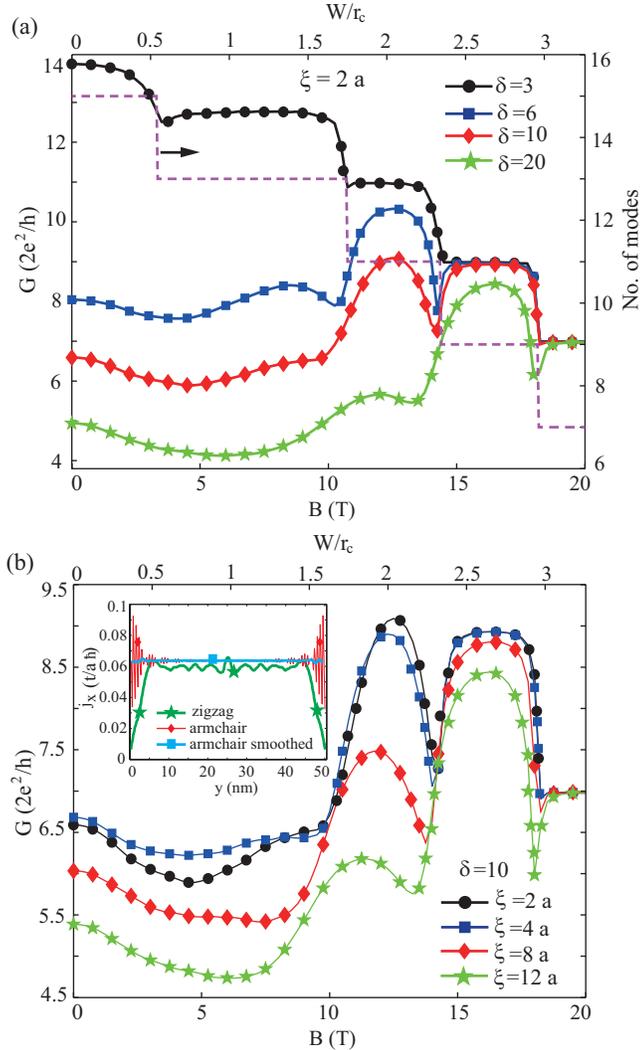}
\caption{(Colour online) Magnetoconductance of a GNR with zigzag edges and otherwise identical parameters to those in Fig. 1, for various edge disorder amplitudes and a fixed correlation length of $\xi=2a$ (a), where the dashed line denotes the number of occupied modes, and for different correlation lengths at a disorder amplitude of $\delta=10$ (b). The inset in (b) shows the cross-sectional current density for zigzag and armchair GNRs with perfect edges.}
\label{ERID_Fig2}
\end{figure}

Fig. \ref{ERID_Fig2} shows the corresponding results for zigzag GNRs. Again, the ERID structures are observed at cyclotron radii somewhat larger than $W$, but distinct differences to armchair GNRs can be identified. First of all, at $B=0$, the conductance suppression by the edge disorder is less pronounced as compared to armchair GNRs. This represents the well-known reduced sensitivity to edge disorder of zigzag GNRs compared to armchair GNRs.\cite{Areshkin2007}. In addition, at small disorder amplitudes, e.g. for $\delta= 3$, uppermost trace in Fig. \ref{ERID_Fig2}(a), the ERID is absent and the magnetoconductance reflects just the depopulation of the modes. As $\delta$ increases to 6, the ERID becomes visible, and its amplitude approaches that one observed in armchair GNRs as $\delta$ is increased. As $\xi$ is increased at an amplitude of $\delta=10$, Fig. \ref{ERID_Fig2} (b), the ERID amplitude develops nonmonotonically and shows a minimum around $\xi=4a$. For larger correlation lengths, the ERID develops an asymmetric shape.

Apparently, compared to armchair GNRs, the magnetoconductance in zigzag GNRs is less sensitive to edge disorder for small disorder amplitudes. In order to interpret this difference, we take a look at the current density distribution $j_x(y)$ in GNRs with perfect edges, see the inset in Fig. \ref{ERID_Fig2} (b).
The current density of armchair GNRs oscillates as a function of the site index (red trace), which is a consequence of the alternating character of the wave functions. Averaging over 4 adjacent atoms (blue trace) however shows that the current density is almost homogeneously distributed over the cross section. In zigzag GNRs, however, the current density is strongly suppressed in an interval extending about $\approx 5\,\mathrm{nm}$ from the edge into the bulk. This current density suppression, which has been reported earlier, \cite{Zarbo2007} suggests that zigzag GNRs are less sensitive to edge disorder, in particular for disorder amplitudes below $\approx 5\,\mathrm{nm}$ for the Fermi energy considered here, which is in good qualitative agreement with the magnetoconductance simulations. Furthermore, it has been shown that in the single-mode regime in zero magnetic field,  the
conductance of the zigzag GNRs remains practically unaffected by the edge
disorder \cite{Areshkin2007}. In this case, the
propagating channels in the zigzag GNRs are edge states flowing on the
opposite boundaries \cite{Wakabayashi1999,Wakabayashi2007} such that the backscattering is
suppressed. We believe that this mechanism is not relevant in our study since we are in the multimode regime where the special character of the edge state should be insignificant. It would thus be interesting to study the magnetoconductance very in close proximity to the charge neutrality point. Our numerical method is an impractical tool for such a study due to the onset of Coulomb blockade effects as the Fermi energy is reduced.\cite{Droescher2011} We did, however, vary the number of occupied subbands at fixed Fermi energy by changing $W$ and found that the ERID remains located close to $r_c=W$.

\begin{figure}[tbp]
\includegraphics[width=85mm]{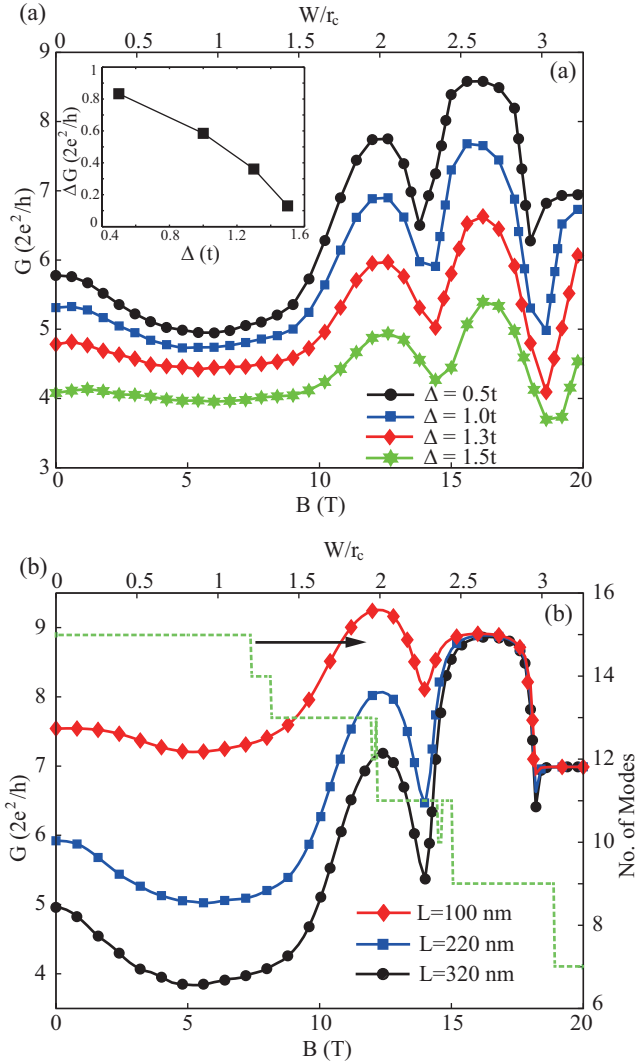}
\caption{(Colour on-line) (a) Effect of bulk disorder on the edge roughness induced
magnetoconductance dip in armchair GNRs, averaged over 500 configurations.
The width and length of GNR are $50\,\mathrm{nm}$ and $220\,\mathrm{nm}$,
respectively. The edge disorder parameters are $\protect\delta=10$ and $%
\protect\xi=2a$, while the bulk disorder is varied by tuning impurity center
height $\Delta$. The inset shows the ERID amplitude as a function of $\Delta$. (b) Length dependence of the magnetoconductance. }
\label{ERID_Fig3}
\end{figure}

To discuss implications regarding the experimental observability of the
ERID, we first study the effect of bulk disorder. The edge disorder of an armchair GNR is fixed to the parameters $\delta=10$ and $\xi=2a$ as used in Fig. \ref{ERID_Fig1}. Fig. \ref{ERID_Fig3} (a) shows the
magnetoconductance for various values of the bulk disorder potential
amplitudes $\Delta$, for short-range disorder ($\Lambda =a$) and with the impurity density $n_{imp}=8\times 10^{12}cm^{-2}$. As $\Delta$ gets larger, the absolute amplitude of the ERID gets
suppressed while its position remains constant. In addition, a small
conductance dip emerges at $B=0$ as the bulk disorder is increased. We
interpret this as the onset of bulk-disorder induced weak localization. This behavior remains the same for long range potentials ($\Lambda=32 a$, not shown).

Second, the length dependence of the ERID is studied. As the length $L$ of the GNR is increased from $100\,\mathrm{nm}$ to $200\,\mathrm{nm}$, the ERID amplitude increases and tends to saturate as $L$ is increased further. We take this as an indication that the system goes from the ballistic into the diffusive regime around a length of $200\,\mathrm{nm}$. We note that in all cases studied here, the localization
length $\ell_{loc}$ is larger than $L$. For the case shown in
Fig. \ref{ERID_Fig3} (b), $\ell_{loc}= 2\mu m$ is
found by fitting the scaling law $\ln(1+1/g)=L/\ell_{loc}$ with $g$ denoting the
dimensionless conductance $g=G/(e^2/\pi\hbar)$. \cite{Xu2009,LeePA1985} For
GNRs with $L>\ell_{loc}$, the conductance would be exponentially suppressed.

These investigations allow us to draw some conclusions regarding the
observability of the ERID in GNRs in experiments and concerning the
suitability of this effect to characterize the edge roughness. It has
emerged that the ERID can be observed in GNRs only if the bulk disorder level is sufficiently low. This may not be the case in present experimental implementations. Typical experimental edge roughness amplitudes are in
the range of 1-10 missing atoms, which is the range where the wire peak
should be observable in armchair GNRs provided the correlation length of the edge roughness
remains small, i.e., of the order of the Fermi wavelength or below. In zigzag GNRs, on the other hand, the ERID will be only visible if the edge roughness amplitude is not small compared to the interval of suppressed current density at the GNR edges, i.e. $\delta \geq 6$ for the parameter range studied here. Like in
conventional quantum wires, scattering at the edges gets less diffusive if
the roughness becomes smoother which suppresses the ERID. Furthermore, the
GNR lengths should be smaller than the localization length but well within the diffusive regime for maximum
visibility. We note that there is no dephasing mechanism in our model. In
real samples with finite dephasing time, the conductance around B=0 will
increase due to the suppression of weak localization.

In summary, we have studied the magnetotransport properties of both armchair
and zigzag GNRs in the presence of edge disorder. In both types of GNRs, an edge
roughness - induced conductance minimum is found at cyclotron radii corresponding
roughly to the GNR width, provided the edge disorder has a correlation
length not much larger than the Fermi wavelength and the bulk disorder is
sufficiently low. The effect is most visible in the diffusive regime. This magnetoconductance structure is similar in character to that one observed at conventional, rough quantum wires. Furthermore, in zigzag GNRs, it gets suppressed as the edge disorder amplitude drops below the spatial extension of the reduced current density.

\acknowledgments
H.X. and T.H. acknowledge financial support from Heinrich-Heine-Universit\"{a}t D\"{u}sseldorf


\begin{thebibliography}{99}
\expandafter\ifx\csname url\endcsname\relax\def\url#1{\texttt{#1}}\fi

\bibitem{Novoselov2005}
\Name{Novoselov K.~S., Geim A.~K., Morozov S.~V., Jiang D., Katsnelson M.~I.
  \and Grigorieva I.~V.} \REVIEW{Nature}{438}{2005}{197}.

\bibitem{CastroNeto2009}
\Name{Neto A. H.~C., Guinea F., Peres N. M.~R., Novoselov K.~S. \and Geim
  A.~K.} \REVIEW{Rev. Mod. Phys.}{81}{2009}{109}.

\bibitem{Morozov2008}
\Name{Morozov S.~V., Novoselov K.~S., Katsnelson M.~I., Schedin F., Elias
  D.~C., Jaszczak J.~A. \and Geim A.~K.} \REVIEW{Phys. Rev. Lett.}{100}{2008}{016602}.

\bibitem{Areshkin2007}
\Name{Areshkin D.~A., Gunlycke D. \and White C.~T.} \REVIEW{Nano Lett.}{7}{2007}{204}.

\bibitem{Evaldsson2008}
\Name{Evaldsson M., Ihnatsenka S. \and Zozoulenko I.~V.} \REVIEW{Phys. Rev. B}{77}{2008}{165306}.

\bibitem{Xu2008}
\Name{Xu H., Heinzel T., Evaldsson M. \and Zozoulenko I.~V.} \REVIEW{Phys. Rev.
  B}{77}{2008}{245401}.

\bibitem{Han2007}
\Name{Han M.~Y., Ozyilmaz B., Zhang Y. \and Kim P.} \REVIEW{Phys. Rev. Lett.}{98}{2007}{206805}.

\bibitem{Peres2011}
\Name{Peres N. M.~R.} \REVIEW{Rev. Mod. Phys.}{82}{2011}{2673}.

\bibitem{Sarma2011}
\Name{Sarma S.~D., Adam S., Hwang E.~H. \and Rosso E.} \REVIEW{Rev. Mod. Phys.
  }{83}{2011}{407}.

\bibitem{Stampfer2009}
\Name{Stampfer C., Guettinger J., Hellmueller S., Molitor F., Ensslin K. \and
  Ihn T.} \REVIEW{Phys. Rev. Lett.}{102}{2009}{056403}.

\bibitem{Liu2009}
\Name{Liu X.~L., Oostinga J.~B., Morpurgo A.~F. \and Vandersypen L. M.~K.}
  \REVIEW{Phys. Rev. B}{80}{2009}{121407}.

\bibitem{Todd2009}
\Name{Todd K., Chou H.~T., Amasha S. \and Goldhaber-Gordon D.} \REVIEW{Nano
  Lett.}{9}{2009}{416}.

\bibitem{Thornton1989}
\Name{Thornton T.~J., Roukes M.~L., Scherer A. \and van~de Gaag B.~P.}
  \REVIEW{Phys. Rev. Lett.}{63}{1989}{2128}.

\bibitem{Akera1991}
\Name{Akera H. \and Ando T.} \REVIEW{Phys. Rev. B}{43}{1991}{11676}.

\bibitem{Lettau1994}
\Name{Lettau C., Wendel M., Schmeller A., Hansen W., Kotthaus J.~P., Klein W.,
  Traenkle G., Weimann G. \and Holland M.} \REVIEW{Phys. Rev. B}{50}{1994}{2432}.

\bibitem{Schapers2006}
\Name{Sch$\rm{\ddot{a}}$pers T., Guzenko V.~A., Pala M.~G.,
  Z$\rm{\ddot{u}}$licke U., Governale M., Knobbe J. \and Hardtdegen H.}
  \REVIEW{Phys. Rev. B }{74}{2006}{081301}.

\bibitem{Gilbertson2011}
\Name{Gilbertson A.~M., Fearn M., Korm\'{a}nyos A., Read D.~E., Lambert C.~J.,
  Emeny M.~T., Ashley T., Solin S.~A. \and Cohen L.~F.} \REVIEW{Phys. Rev. B
  }{83}{2011}{075304}.

\bibitem{Poumirol2010}
\Name{Poumirol J.-M., Cresti A., Roche S., Escoffier W., Goiran M., Wang X., Li
  X., Dai H. \and Raquet B.} \REVIEW{Phys. Rev. B}{82}{2010}{041413}.

\bibitem{Bai2010}
\Name{Bai J., Cheng R., Xiu F., Liao L., Weng M., Shailos A., Wang K.~L., Huang
  Y. \and Duan X.} \REVIEW{Nature Nanotech.}{9}{2010}{655}.

\bibitem{Oostinga2010}
\Name{Oostinga J.~B., Sacepe B., Craciun M.~F. \and Morpurgo A.~F.}
  \REVIEW{Phys. Rev. B}{81}{2010}{193408}.

\bibitem{Ribeiro2011}
\Name{Ribeiro R., Poumirol J.-M., Cresti A., Escoffier W., Goiran M., broto
  J.-M., Roche S. \and Raquet B.} \REVIEW{arXiv:1102.3300v1[cond-mat.mes.hall]
  }{}{2011}{}.

\bibitem{Adam2009}
\Name{Adam S., Brouwer P.~W. \and Sarma S.~D.} \REVIEW{Phys. Rev. B}{79}{2009}{201404}.

\bibitem{Klos2009}
\Name{Klos J.~W., Shylau A.~A., Zozoulenko I.~V., Xu H. \and Heinzel T.}
  \REVIEW{Phys. Rev. B}{80}{2009}{245432}.

\bibitem{Rycerz2007}
\Name{Rycerz A., Tworzydlo J. \and Beenakker C. W.~J.} \REVIEW{Europhys.
  Lett.}{79}{2007}{57003}.

\bibitem{Lewenkopf2008}
\Name{Lewenkopf C.~H., Mucciolo E.~R. \and Castro~Neto A.~H.} \REVIEW{Phys.
  Rev. B}{77}{2008}{081410}.

\bibitem{Zarbo2007}
\Name{Zarbo L.~P. \and Nikolic B.~K.} \REVIEW{Europhys.
  Lett.}{80}{2007}{47001}.

\bibitem{Droescher2011}
\Name{Dr$\rm{\ddot{o}}$scher S., Knowles H., Meir Y., Ensslin K, \and Ihn T. \and Nikolic B.~K.} \REVIEW{Phys. Rev. B}{84}{2011}{073405}.

\bibitem{Wakabayashi1999}
\Name{Wakabayashi K., Fujita M., Ajiki H. \and Sigrist M.} \REVIEW{Phys. Rev. B}{59}{1999}{8271}.

\bibitem{Wakabayashi2007}
\Name{Wakabayashi K., Takane Y. \and Sigrist M.} \REVIEW{Phys. Rev. Lett.}{99}{2007}{036601}.

\bibitem{Bardarson2007}
\Name{Bardarson J.~H., Tworzydlo J., Brouwer P.~W., \and Beenakker C. W.~J.} \REVIEW{Phys. Rev. Lett.}{99}{2007}{106801}.

\bibitem{Xu2009}
\Name{Xu H., Heinzel T. \and Zozoulenko I.~V.} \REVIEW{Phys. Rev. B}{80}{2009}{045308}.

\bibitem{LeePA1985}
\Name{Lee P.~A. \and Ramakrishnan T.~V.} \REVIEW{Rev. Mod. Phys.}{57}{1985}{287}.

\end{thebibliography}
\end{document}